\documentstyle[aps]{revtex}

\begin{document}
\draft

\title{$SO(10)$ GUTs with Gauge Mediated Supersymmetry Breaking}
\author{M.~Frank$^{\text a}$, H.~Hamidian$^{\text b}$, and
K.~Puolam\"{a}ki$^{\text c}$}
\address{
$^{\text a}$Department of Physics, Concordia University, 1455 De
Maisonneuve Blvd. W., Montr\'{e}al, PQ H3G~1M8, Canada}
\address{
$^{\text b}$Department of Physics, University of Illinois at Chicago, 845
W. Taylor Street, Chicago, IL 60607, USA}
\address{
$^{\text c}$Helsinki Institute of Physics, P.O.Box 9, FIN-00014 University
of Helsinki, Finland}
\date{March 30, 1999}
\maketitle

\begin{abstract}
We explore the phenomenology of supersymmetric $SO(10)$ grand unified
theories with gauge mediated supersymmetry breaking. We show that if
$SO(10)$ breaking proceeds through intermediate left-right symmetric
gauge groups which are broken at the supersymmetry breaking scale,
then perturbative unification allows the existence of only a few
consistent models with very similar phenomenological consequences. We
list and discuss some distinctive signatures of these theories. The
most remarkable feature of the class of theories introduced here is that,
unlike in models with simpler symmetry breaking chains,
the set of allowed messengers is practically unique.
\end{abstract}


{\it Introduction}. If the standard model (SM) indeed descends from a
supersymmetric theory, a number of criteria must be met and some major
questions need to be answered. The most important question among these
is to understand the origin and nature of supersymmetry (SUSY)
breaking.  The currently popular models include gravity-mediated SUSY
breaking, gauge-mediated SUSY breaking, and $U(1)$-mediated SUSY
breaking
\cite{mohapatratasi}.
Each of these models exhibits certain desirable features and drawbacks
of its own; however, the main underlying assumption, i.e. the existence of
a hidden sector where SUSY is broken and the means by which the breaking is
transmitted to the visible (low-energy) world of the SM,
is shared by all of these approaches. In theories with gauge mediated SUSY
breaking
(GMSB) supersymmetry-breaking soft terms are generated by a set of
particles, called messengers, at a scale $\Lambda_M$ (called the
messenger scale) which is {\it a priori} unrelated to the GUT scale
\cite{dineetal}.
A distinguishing and attractive feature of GMSB models is that they
naturally lead to degenerate squark and slepton masses and thus
alleviate the flavor changing neutral current (FCNC) problem of
the minimal supersymmetric standard model (MSSM). Furthermore, GMSB theories
are highly predictive; they lead to a dramatic reduction of the number of free
parameters and their predictions will be testable in the not-too-distant
future
\cite{gr}.

In GMSB theories SUSY breaking in the (unspecified) hidden sector is
communicated to the visible sector through the
$SU(3)_C \times SU(2)_L \times U(1)_Y$ SM gauge interactions of the
``messenger''
fields with the visible sector. In the minimal version of GMSB, the
messenger
fields belong to the ${\bf 5}+{\overline {\bf 5}}$ or ${\bf 10}+{\overline
{\bf 10}}$
representations of the $SU(5)$ gauge group and there exists at least one
singlet
superfield $S$ which couples to vector-like messenger superfields
$V + {\overline V}$
through the superpotential interaction
\begin{equation}
W_{mess} = \lambda_V S V {\overline V},
\label{wmess}
\end{equation}
where the Yukawa couplings, $\lambda_V$'s in (\ref{wmess}),
are assumed to coincide at the unification scale $M_G$.
Consequently, the spectrum at the messenger scale consists of a set of
fields in complete $SU(5)$ representations and the mass splitting among
the fields can be determined from the renormalization group (RG)
running of
the messenger Yukawa couplings from $M_G$ down to the messenger
scale $\Lambda_M$.
The (generalized) non-minimal versions of GMSB theories,
in which the messenger fields do not necessarily form complete $SU(5)$ GUT
multiplets, have also been studied by a number of authors
\cite{martin,hamidian,hhpz}.
These studies indicate that GMSB theories based on $SU(5)$ GUTs
are phenomenologically disfavored. However,
since the high predictability and simplicity of GMSB theories are hard to
achieve otherwise, it is important to study other (SUSY) GUT gauge groups
which
may lead to more realistic models.

An attempt to go beyond the minimal GMSB gauge group was made in
\cite{mohapatra}
where the
authors embedded the electroweak gauge group in the gauge group
$SU(2)_L \times U(1)_{I_{3R}} \times U(1)_{B-L}$
[or $SU(2)_L \times SU(2)_R \times U(1)_{B-L}$] at the SUSY breaking scale
$\Lambda_{SUSY}$.
The distinguishing features of these models include the automatic
conservation
of $R$-parity, non-vanishing neutrino masses, and a unified
hidden-plus-messenger sector
potential.
This has led us to investigate whether
$SO(10)$ SUSY GUTs (which contain these gauge groups) with
GMSB could lead to phenomenologically attractive scenarios.
It is the purpose of this Letter to show that not only does this seem to be the
case,
but also that there are only a few scenarios (with very similar testable
predictions)
which are consistent with all the constraints.

{\it $SO(10)$ Models and GMSB}. The usual model building
in GMSB theories based on the $SU(5)$ gauge group suffers from a number of
serious
drawbacks and leaves some important questions unanswered.
In particular, one encounters problems related to the nucleon decay rates,
lack of a natural
mechanism to generate neutrino masses,
the existence of arbitrary $R$-parity violating interactions, and the SUSY
$CP$
problem.
A viable alternative for naturally avoiding most of these problems is
to consider $SO(10)$ grand unification instead. In $SO(10)$ GUTs the dangerous
colour triplet Higgs boson can naturally be made very heavy---thus
suppressing rapid proton decay ---and by incorporating left-right (LR)
supersymmetric
theories, which can arise through $SO(10)$ breaking, the SUSY $CP$
problem can be solved while automatically conserving $R$-parity
\cite{mohras}.
Furthermore, as an
additional bonus, since a full generation of fermions comes in one spinorial
representation which includes right-handed neutrinos,
the see-saw mechanism can be used to naturally generate small neutrino
masses \cite{bb}.

$SO(10)$ symmetry breaking can proceed in essentially two ways:
$SO(10)$ can break down to $SU(5) \times U(1)$ at $\sim 10^{18}$~GeV
with a further breaking down to the MSSM at $\sim 10^{16}$~GeV.
Alternatively, $SO(10)$ can break down to some
LR symmetric gauge group $G_{LR}$ (such as $G_{224}=SU(2)_L\times SU(2)_R\times
SO(4)_C$) with a subsequent breaking down to the MSSM at some
intermediate scale.
Here we assume that $SO(10)$ first breaks down to an intermediate LR
symmetric group $G_{LR}$ at the scale
$M_G$, followed by the breaking of $G_{LR}$ at the scale $M_R$ down to the
MSSM. The scale
$M_G$ lies below the Planck scale  $\sim 10^{19}$~GeV and must be no less
than $10^{16}$~GeV to ensure nucleon stability.
To keep things as simple as possible, we further
assume that there are no other symmetry breaking scales between $M_R$ and
$M_G$.

The supersymmetric ($SO(10)$-based) LR models described here have the
gauge groups: $G_{LR}^I=SU(2)_L \times U(1)_{I_{3 R}} \times SU(3)_C
\times U(1)_{B-L}$ and $G_{LR}^{II}=SU(2)_L \times SU(2)_R \times
SU(3)_C \times U(1)_{B-L}$, and further assumptions are needed to
include SUSY breaking in these models.  Since chirality plays a very
important role in SUSY it is very natural to expect SUSY and LR
symmetry breaking scales to be somehow related.  A very attractive
possibility is to simply take $\Lambda_{SUSY} = M_R$ which not only
connects the SUSY breaking and the gauge symmetry breaking scales, but
also requires that the electroweak symmetry breaking remain radiative.
A consequence of this assumption is that $\Lambda_{SUSY} \sim M_R \sim
100$~TeV, which follows from the usual (MSSM) requirement that the
sparticle masses stay in the TeV range.  We will thus limit the ranges
of the left-right breaking and the GUT scales as:
\begin{equation}
10^5~{\text{GeV}}<M_R<10^7~{\text{GeV}} {\text{ and }}
10^{16}~{\text{GeV}}<M_G<10^{19}~{\text{GeV}}.
\end{equation}
Furthermore, as is usually done in the study of SUSY GUTs, we assume
that the gauge couplings unify at $M_G$ and remain perturbative up to
the Planck scale.  Putting all these ingredients together we now
proceed by specifying the LR models studied here.

{\it Model~I:}~$SU(2)_L\times U(1)_{I_{3R}} \times U(1)_{B-L}$. This
model is phenomenologically interesting for a number of reasons: It
contains all the usual matter multiplets {\em plus} right-handed
neutrinos and (due to the absence of baryon-number violating terms in
the superpotential) automatically conserves $R$-parity.  The
superpotential which describes this model is:
\begin{eqnarray}
\label{superpotential1}
W & = &  h_{u} Q H_u u^{c} + h_d Q H_d d^{c} + h_e L H_d e^{c}
+ h_{\nu} L H_u \nu^{c} \nonumber \\
& & + \mu H_u H_d + f \delta \nu^c \nu^c + M_{R} \delta {\bar \delta} 
+W_m,
\label{model1w}
\end{eqnarray}
where $W_m$ is the messenger sector superpotential.
The particle content of this model consists of: The doublets
$Q(2,0,1/6)$, $L(2,0,-1/2)$, the singlets ~$u^c(1,-1/2, -1/6)$,~$d^c(1, 1/2,
-1/6)$, $e^c(1, 1/2, 1/2)$, $\nu^c(1, -1/2, 1/2)$,
the Higgs doublets
$H_u(2, 1/2, 0)$ and
$H_d( 2, -1/2, 0)$, which are the same as in the MSSM, two Higgs triplets,
$\delta
(1, 1, -1)$ and ${\bar \delta} (1, -1, 1)$, which break the
$U(1)_{I_{3R}} \times U(1)_{B-L}$ symmetry down to $U(1)_Y$ of the
standard model, the gauge bosons $W(3, 0, 0)$,~$B(1, 0, 0)$, and $V(1, 0,
0)$, and the superpartners of all these fields.

{\it Model~II:}~$SU(2)_L\times SU(2)_R \times U(1)_{B-L}$. The
phenomenology of the LR supersymmetric model based on the
gauge group $SU(2)_L\times SU(2)_R \times U(1)_{B-L}$ has been extensively
studied in recent years
\cite{fh}.
Among other attractive features, this model simultaneously solves
both the strong and weak $CP$ problems and can also accomodate automatic
conservation of $R$-parity.
This model can be specified by the superpotential:
\begin{eqnarray}
\label{superpotential2}
W & = & {\bf h}_{q}^{(i)} Q_L^T\tau_{2}\Phi_{i} \tau_{2}Q_R +
{\bf h}_{l}^{(i)}
L_L^T\tau_{2}\Phi_{i} \tau_{2}L_R \nonumber \\
& &+ i({\bf h}_{LR}L_L^T\tau_{2} \delta_L L_L + {\bf
h}_{LR}L_R^{T}\tau_{2}
\Delta_R L_R) \nonumber \\ & & + M_{LR}\left[ {\rm Tr} (\Delta_L
{\delta}_L + {\rm Tr} (\Delta_R {\delta_R}) \right]
\nonumber \\ & &+ \mu_{ij} {\rm Tr} (\tau_{2}\Phi^{T}_{i} \tau_{2}
\Phi_{j}) + W_m,
\label{model2w}
\end{eqnarray}
where $W_m$ denotes the messenger sector superpotential. The particle
content of this model consists of the doublets $Q_L(2,1,1/6)$,
$L_L(2,1,-1/2)$, $ Q_R(1, 2, -1/6)$, $L_R( 1, 2 ,1/2)$, the bi-doublet
Higgs fields $\Phi_u (2, 2, 0)$ and $\Phi_d( 2, 2, 0)$, the triplet
Higgs fields $\Delta_L (3,1,-1)$, $\Delta_R (1,3,-1)$, $\delta_L
(3,1,1)$, and $\delta_R (1,3,1)$, the gauge bosons $W_L(3, 1, 0)$,
$W_R(1, 3, 0)$, and $V(1, 1, 0)$, and the corresponding superpartners.

In both models I and II the relation between the $U(1)_{B-L}$ gauge
coupling $\alpha_{B-L}$ and the GUT-normalized gauge coupling
$\alpha_V$ is fixed through $\alpha_V= \frac 23 \alpha_{B-L}$.  The
messenger sector in both models is described by $N_f$ flavours of
chiral superfields $\Phi_i$ and ${\overline \Phi}_i$ $(i=1,\cdots,
N-f)$ which belong to the ${\bf r} +{\bf {\overline r}}$
representation of the corresponding gauge group.

If one assumes gauge coupling constant unification and requires that the
messengers form complete GUT multiplets, then the
presence of intermediate scale messenger fields leaves $M_G$ unchanged. In
this case, due to the
contribution of the messenger fields, one obtains \cite{dineetal}:
\begin{eqnarray}
\delta \alpha^{-1}_{GUT}=-\frac{N}{2 \pi}\ln\frac{M_G}{\Lambda_M},
\end{eqnarray}
where
$N=\sum^{N_f}_{i=1} n_i$, with
$n_i$ denoting twice the Dynkin index of the ${\bf r}$ representation for
the $i$th flavor. This leads to
the constraint:
\begin{eqnarray}
N \alt 150/\ln \frac{M_G}{\Lambda_M},
\end{eqnarray}
as a result of the perturbativity hypothesis at the GUT scale.

To search for possible GMSB scenarios that are consistent with the
assumptions and constraints that we have
invoked so far, let us begin by listing the
messenger fields. In choosing messenger fields, which form the
messenger
 sector, we maintain the constraint that they should occupy the same
representation as Model~I or Model~II chiral superfields. (The motivation
is that stable particles with exotic  quantum
numbers are a disaster for cosmology \cite{martin}). The possible
messenger fields in Model~I, which transform under the gauge group
$SU(3)_C \times SU(2)_L \times U(1)_{I_{3 R}} \times U(1)_{B-L}$, are given by:
$
Q_8 = (8, 1, 0, 0),
L_3 = (1, 3, 0, 0),
\Delta+{\overline \Delta} = (1, 3 , 0, -1) +conj.,
\Delta^c +{\overline {\Delta ^c}} = (1, 1 , -1, 1) +conj.,
H + {\overline H} =  (1, 2, \frac{1}{2}, 0) + conj.,
Q + {\overline Q} = (3, 2, 0, \frac{1}{6}) + conj.,
U ^c+{\overline {U^c}} =({\overline 3}, 1, - \frac{1}{2}, -\frac{1}{6})
+conj.,
D ^c+{\overline {D^c}}=({\overline 3}, 1 ,\frac{1}{2},- \frac{1}{6})
+conj.,
L + {\overline L} = (1, 2, 0,  -\frac{1}{2}) +conj.,
e^c +{\overline {e^c}} = (1, 1, \frac {1}{2} , \frac{1}{2}) +conj.,
\nu^c +{\overline {\nu^c}}= ( 1, 1, -\frac {1}{2}, \frac{1}{2}) +conj.,
$
and those of Model~II, transforming under the  $SU(3)_C\times SU(2)_L
\times SU(2)_R \times U(1)_{B-L}$
gauge group, are given by:
$
Q_8=(8,1,1,0),
Q_{3}=(1,3,1,0),
Q_{3}^c=(1,1,3,0),
\phi=(1,2,2,0),
Q +{\overline Q}=(3,2,1,\frac 16)+conj.,
Q^c +{\overline {Q^c}}=(3,1,2,-\frac 16)+conj.,
L+{\overline L}=(1,2,1,-\frac 12)+conj.,
L^c+{\overline {L^c}}=(1,1,2,\frac 12)+conj.,
\Delta+{\overline \Delta} =(1,3,1,-1)+conj.,
\Delta^c+{\overline {\Delta^c}} =(1,1,3,1) +conj.
$.

To restrict the messenger sector by using the unification requirement the
RG $\beta$-functions and
the matching conditions must be used.
At the LR breaking scale $M_R$
the couplings are required to match those of the MSSM.  Denoting the
$\beta$-functions of the $SU(2)_L$, $SU(2)_R$, $U(1)_{B-L}$,
and $SU(3)_C$ gauge groups respectively by $\beta_L$, $\beta_R$,
$\beta_{V}$, and $\beta_C$, the one-loop RG equations at the
scale $M_R$ are:
\begin{eqnarray}
\alpha_I^{-1}(M_R) = \alpha_G^{-1} +\beta_I (t_G-t_R),
\label{eq:rng1}
\end{eqnarray}
where $t_R = \frac 1{2 \pi} \ln \frac{M_R}{M_Z}$, $t_G = \frac 1{2 \pi} \ln
\frac{M_G}{M_Z}$, and $I=L,R,V,C$.

The one-loop matching conditions at the scale $M_R$ read as follows:
\begin{eqnarray}
\frac {5}{3} \alpha_1^{-1}(M_R)& &=\alpha_R^{-1}(M_R)+\frac {2}{3}
\alpha_{V}^{-1}(M_R) , \nonumber \\
\alpha_2^{-1}(M_R)& &=\alpha_L^{-1}(M_R) , \nonumber \\
\alpha_3^{-1}(M_R)& &=\alpha_C^{-1}(M_R),
\label{eq:rng2}
\end{eqnarray}
where $\alpha_1$, $\alpha_2$, and $\alpha_3$ correspond to the
$U(1)_Y$, $SU(2)_L$, and $SU(3)_C$ gauge couplings respectively.
Combining equations (\ref{eq:rng1}) and (\ref{eq:rng2}) then yields:
\begin{equation}
\alpha_k^{-1}(M_Z)=\alpha_G^{-1}+\beta_k^{MSSM}(t_R) + \beta_k^{LR}
(t_G-t_R),
\label{eq:rng3}
\end{equation}
with $k=1,2,3$ and
\begin{equation}
\beta^{LR}=\left( \begin{array}{c} \beta_1^{LR} \\
\beta_2^{LR} \\ \beta_3^{LR}
\end{array} \right) =\left(  \begin{array}{c}
\frac 35 \beta_R+\frac
25 \beta_{V} \\ \beta_L \\ \beta_C \end{array}
\right),
\label{eq:definebeta}
\end{equation}
where the MSSM $\beta$-functions $\beta_k^{MSSM}=(33/5,1,-3)^T$ have been
used.

By eliminating $\alpha_G$ from equation (\ref{eq:rng3})
it is straightforward to obtain the
following limits on the differences between the LR $\beta$
functions:
\begin{equation}
3.1 < \beta_2^{LR}-\beta_3^{LR} < 4.1 {\text{
and }} 7.4 <
\beta_1^{LR}-\beta_3^{LR} < 9.9,
\label{eq:constrain1}
\end{equation}
and demanding that the gauge couplings remain perturbative all the way up
to the Planck scale
equation (\ref{eq:rng3}) yields:
\begin{equation}
\beta_1^{LR}<10.4 {\text{ , }} \beta_2^{LR}<6.1
{\text{ and }}
\beta_3^{LR}<3.0 ,
\label{eq:constrain2}
\end{equation}
which constrains the number of messenger fields.

By examining the $\beta$ functions for
Models I and II along with the constraints (\ref{eq:constrain1}) and
(\ref{eq:constrain2}) we find that there are
no consistent solutions in
Model~II. Model~I, on the other hand, leads to consistent solutions with
the messenger multiplicities:
\begin{equation}
n_8=n_3=n_H+n_L=1 {\text{ and }}
n_{e^c}+n_{\nu^c}=0,1 .
\label{eq:messconstrain}
\end{equation}

According to (\ref{eq:messconstrain}) the messenger sector in Model~I
consists of a color octet ($n_8=1$),
an $SU(2)_L$ triplet ($n_3=1$),
a pair of $H$
or $L$ type messenger fields
($n_H+n_L=1$), and a pair of
$e^c$ or $\nu^c$ type fields
($n_{e^c}+n_{\nu^c}=0,1$). There are thus a
total of six solutions for the
messenger multiplicities, all of which (as will be described in more detail
elsewhere
\cite{fhp2})
lead to very similar mass
parameters for the MSSM and render our scheme
extremely predictive. In all these cases the $SU(2)_L$ and
$SU(3)_C$
gauge couplings meet at $M_G \simeq 2.0 \times 10^{16}$~GeV
for $10^5~{\text{GeV}} < M_R < 10^7~{\text{GeV}}$.
For the solution with $n_{e^c}+n_{\nu^c}=0$, $\alpha_1$
and
$\alpha_3$ meet at $M_G' \simeq 1.9
\times 10^{16}$~GeV, which is within 6\% of the GUT scale. When
$n_{e^c}+n_{\nu^c}=1$
the mismatch is much worse; in this case $\alpha_1$ and $\alpha_3$
meet at $M_G' \simeq 4-5 \times 10^{15}$~GeV. However, since the
mismatch can be attributed to threshold effects, we will consider
this solution as well. Another constraint on the models studied here
comes from the ratio
$
\tan^2
\theta_R=\alpha_{B-L}(M_R)/{\alpha_R}(M_R),
$ which is a very important parameter in LR theories. It is completely
determined by the unification condition and the chosen messenger
multiplicities; in our case one has $1.3 \le \tan^2
\theta_R \le 1.6$, as shown in table~\ref{tab:models}. To compute the
sparticle mass spectrum for the
models studied here this value must be taken into account.

{\it The sparticle spectrum}.
The messenger masses in the LR models described here can be calculated once
the messenger sector and the messenger scale are fixed.
By matching the LR model and the MSSM the values of the MSSM parameters
at the messenger scale and the RG $\beta$-functions can be used to
calculate the full MSSM
particle spectrum as a function of $\tan\beta$
\cite{hamidian}.
One could further reduce the number of parameters
and solve the supersymmetric CP-problem, as was done
in
\cite{hhpz},
by requiring that the bilinear scalar coupling $B$ vanish at the
messenger scale; a condition which fixes $\tan\beta$.  Here, however,
we wish to consider a broader possibility by letting $\tan\beta$ (or
equivalently the bilinear scalar coupling) remain a free
parameter. The resulting sparticle spectrum---calculated in terms of
$\Lambda_{SUSY}, \Lambda_M$, and $\tan \beta$---can generally be
divided into a light and a heavy sector.  In our case the light sector
consists of the lightest neutralino (${\tilde \chi}^0_1$), the
chargino (${\tilde \chi}^\pm_1$), and the light slepton mass
eigenstates ($\tilde e_1$, $\tilde \mu_1$, $\tilde \tau_1$), while the
squarks comprise the heavy sector.

An important issue which requires particular care when calculating the
sparticle spectrum
is radiative gauge symmetry breaking.
To ensure radiative (gauge) symmetry
breaking, one must check that the resulting vacuum is physical. This can be
achieved
if all the mass-squared eigenvalues of the charged scalars remain positive
and above the
current experimental limits. We have taken these important constraints into
account in
our calculations of the sparticle spectrum (a complete listing of which
will be given in
\cite{fhp2}).
Since the heavy top squark drives
the radiative symmetry breaking, we find that for all values of
parameters the mass-squared term of the up-type Higgs boson acquires a
negative value and thus always leads to radiative symmetry breaking.
[As is usual in GMSB models, the lightest sparticle turns out to
be the lighter stau mass eigenstate and if one
includes the latest LEP2 constraint, $m_{\tilde \tau} \ge 72$~GeV
\cite{LEP2},
then the solutions with squarks lighter than $600$~GeV are immediately
ruled out, as well as the model $n_3=n_8=n_H=1$
and $n_L=n_{e^c}=n_{\nu^c}=0$ with squarks lighter than $1.1$~TeV.]

{\it Conclusions.}  A notable, and somewhat remarkable, aspect of the
GMSB models studied here is that although in general there are six
different consistent solutions, all of them lead to similar
predictions for the supersymmetric mass spectrum (see
table~\ref{tab:models}). The mass spectra for the supersymmetric
partners (and $H^{\pm}$) exhibit certain characteristic features which
we will now summarize: {\it (i)} Depending on the exact messenger
content, the LSP (ignoring the possibility of light gravitino for the
moment) can be either the lighter stau or the lightest neutralino.
Our calculations indicate that the solution with $n_{e^c}=1$ favors
the stau as the LSP, while the one with $n_{e^c}=0$ favors neutralino
as the LSP. Our mass spectrum is dictated by the constraint of keeping
$m^2_{\tilde \tau}$ positive and sufficiently large.  {\it (ii)} As
expected, the lighter selectron is always heavier than the lighter
stau, and sleptons are always lighter than squarks, which turn out to
be very heavy in this model
\cite{fhp2}, always larger than $0.6$ ($1.5$)~TeV for low (high)
$\tan \beta$ solutions. [Note that the
usual mass hierarchy $m_{{\tilde e}_1,2} \leq m_{{\tilde d}_{1,2}}
\approx m_{{\tilde u}_{1,2}} \leq m_{{\tilde t}_{1,2}}$ among the masses,
$m_{{\tilde l,q}_{1,2}}$, of the mixed left and right sleptons and
sqaurks also holds here.]  {\it (iii)} Unlike in supersymmetric models
without GMSB, here the sneutrinos always turn out to be heavier than
the lighter of the charged sleptons with $m_{{\tilde \nu}_{e,\tau}}
\approx m_{{\tilde \tau}_2}$.  {\it (iv)} We find that the bilinear
Higgs coupling $\mu$ lies in the $400-500$~GeV range (for squark
masses of order 1 TeV) and can be positive {\em or} negative, unlike
in \cite{nandi}.  For a vanishing $B$-parameter at the messenger scale
$\mu$ would be positive. In general, the sign of $\mu$ does not seem
to make much difference in the calculated mass spectra. However, there
can be sizable effects is in, e.g., the $b
\rightarrow s\gamma$ decay width, since for positive $\mu$ the
interference between the SM and the chargino contributions is
destructive, whereas for negative $\mu$ it is constructive.  {\it (v)}
The heavy spartner masses in these models turn out to be quite accurately
directly
proportional to the scale $\Lambda_{SUSY}=F/S$.
[As a typical example, for the
choice of $\tan \beta=15$ and $\Lambda_{SUSY}=50$~TeV the
squark masses are around $1$~TeV, while the charged Higgs boson mass is
about $0.5$~TeV in all cases. The heavy sleptons, neutralinos, and
charginos all have masses in the range $310$-$470$~GeV. All of these
heavy masses are linearly proportional to the scale $\Lambda_{SUSY}$.]
We have computed the complete sparticle mass spectrum and the precise
corresponding numerical values for all the six solutions found here
will be reported in
\cite{fhp2}.

{\it Acknowledgements}.  The work of M.F. was partially funded by
NSERC of Canada (SAP0105354).  The work of H.H. is supported by the
U.S. Department of Energy under Grant Number DE-FG02-91ER40676 and
that of K.P. partially by the Academy of Finland (No.~37599).


\begin{table}
\caption{Particle spectrum for a representative set of models with all
possible messenger multiplicities. ($\Lambda_{\text{SUSY}}=50$TeV,
$\Lambda_M=10\Lambda_{\text{SUSY}}$, $\tan\beta=15$, sign$(\mu)=+1$)}
\label{tab:models}
\begin{tabular}{lrrrrrrr}
$(n_3,n_8,n_H,n_L,n_ {e^c},n_{\nu^c})$ & $\tan \beta$ & $\mu$ &
$m_{H^\pm}$ & $m_{{\tilde{\chi}}^\pm_{1,2}}$ &
$m_{{\tilde{\chi}}^0_{1,2,3,4}}$ & $m_{{\tilde{e}}_{1,2}}$ &
$m_{{\tilde{\tau}}_{1,2}}$ \\ $\tan
^2\theta_R=\alpha_{B-L}(M_R)/{\alpha_R}(M_R)$ & $\frac{\Gamma (b
\rightarrow s \gamma)}{\Gamma_{SM}}$ & $M_3$
&
$m_{{\tilde{\nu}}_e}/m_{{\tilde{\nu}}_{\tau}}$ & $m_{{\tilde{u}}_{1,2}}$
&
$m_{{\tilde{t}}_{1,2}}$ & $m_{{\tilde{d}}_{1,2}}$
&
$m_{{\tilde{b}}_{1,2}}$ \\
\tableline $(1,1,1,0,0,0)$ & $15$ & $439$ &
 $514$ &
 $346/469$ &
 $40/346/426/469$ & $70/312$ &
 $54/314$ \\
$1.6$ & $1.2$ & $1060$ & $302/302$ & $1000/1045$ &
 $905/1019$ & $1001/1048$ & $990/1006$ \\
\tableline $(1,1,0,1,0,0)$ & $15$ & $441$ &
 $513$ &
 $347/471$ &
 $40/347/429/470$ & $90/322$ &
 $78/323$ \\
$1.4$ & $1.2$ & $1060$ & $312/312$ & $999/1045$ &
 $904/1019$ & $1000/1048$ & $989/1006$ \\
\tableline $(1,1,1,0,1,0)$ & $15$ & $438$ &
 $514$ &
 $345/468$ &
 $122/346/425/468$ & $104/318$ &
 $93/320$ \\
$1.6$ & $1.2$ & $1060$ & $309/308$ & $1001/1045$ &
 $906/1019$ & $1002/1048$ & $990/1007$ \\
\tableline $(1,1,0,1,1,0)$ & $15$ & $440$ &
 $514$ &
 $346/470$ &
 $122/347/427/470$ & $115/327$ &
 $106/328$ \\
$1.3$ & $1.2$ & $1060$ & $317/317$ & $1000/1045$ &
 $905/1019$ & $1001/1048$ & $990/1007$ \\
\tableline $(1,1,1,0,0,1)$ & $15$ & $438$ &
 $514$ &
 $345/468$ &
 $40/346/425/468$ & $100/318$ &
 $89/320$ \\
$1.6$ & $1.2$ & $1060$ & $308/308$ & $1001/1045$ &
 $906/1019$ & $1002/1048$ & $990/1007$ \\
\tableline $(1,1,0,1,0,1)$ & $15$ & $440$ &
 $513$ & $346/470$ & $40/346/428/470$ & $112/326$ & $103/328$ \\ $1.3$
 & $1.2$ & $1060$ & $317/316$ & $1000/1045$ & $905/1019$ & $1001/1048$
 & $990/1007$ \\
\end{tabular}
\end{table}


\end{document}